# Revisiting Language Support for Generic Programming: When Genericity Is a Core Design Goal

Benjamin Chetioui[a], Jaakko Järvi[b], and Magne Haveraaen[a]

a   University of Bergen, Norway
b   University of Turku, Finland

**Abstract**

**Context**   Generic programming, as defined by Stepanov, is a methodology for writing efficient and reusable algorithms by considering only the required properties of their underlying data types and operations. Generic programming has proven to be an effective means of constructing libraries of reusable software components in languages that support it. Generics-related language design choices play a major role in how conducive generic programming is in practice.

**Inquiry**   Several mainstream programming languages (e.g. Java and C++) were first created without generics; features to support generic programming were added later, gradually. Much of the existing literature on supporting generic programming focuses thus on retrofitting generic programming into existing languages and identifying related implementation challenges. Is the programming experience significantly better, or different when programming with a language designed for generic programming without limitations from prior language design choices?

**Approach**   We examine Magnolia, a language designed to embody generic programming. Magnolia is representative of an approach to language design rooted in algebraic specifications. We repeat a well-known experiment, where we put Magnolia's generic programming facilities under scrutiny by implementing a subset of the Boost Graph Library, and reflect on our development experience.

**Knowledge**   We discover that the idioms identified as key features for supporting Stepanov-style generic programming in the previous studies and work on the topic do not tell a full story. We clarify which of them are more of a means to an end, rather than fundamental features for supporting generic programming. Based on the development experience with Magnolia, we identify variadics as an additional key feature for generic programming and point out limitations and challenges of genericity by property.

**Grounding**   Our work uses a well-known framework for evaluating the generic programming facilities of a language from the literature to evaluate the algebraic approach through Magnolia, and we draw comparisons with well-known programming languages.

**Importance**   This work gives a fresh perspective on generic programming, and clarifies what are fundamental language properties and their trade-offs when considering supporting Stepanov-style generic programming. The understanding of how to set the ground for generic programming will inform future language design.

# The Art, Science, and Engineering of Programming

**Revisiting Language Support for Generic Programming**

## 1 Introduction

It is routine in programming to parameterize algorithms and data structures by type to make them reusable in different contexts. The mechanisms for implementing generic code, however, vary from one language to the other. These details matter: Garcia et al. [30] evaluated and compared the level of support for generic programming in several programming languages (C++, SML, OCaml, Haskell, Eiffel, Java, C#, and Cecil), and showed that many language design choices related to generics significantly influence how conducive that language is in practice to generic programming. This work has had an influence on the design of programming languages (see, e.g., C++'s Concepts [42], Haskell's associated types [14], and Siek and Lumsdaine's idealized 𝒢 language [84, 86] for generic programming).

Generic features are now common features of most widely used languages, and for many of them, these features were an afterthought. The list of such languages has kept growing—examples of languages with recent or planned generic features include Fortran [48], Go [43], ECMAScript, TypeScript, and FlowType. Retrofitting tends to lead to compromises, which raises the questions whether the set of features for generic programming would look the same for languages that incorporate support for generic programming as part of their initial design, and how such potentially different designs support generic programming in practice. This paper sheds light on these questions by examining language designs rooted in algebraic specification. In particular, we conduct a case study and analyse in details the features and programmability of one language representative of the approach, and discuss the findings in the general context of languages that follow the same algebraic design principles. The goal is to inform future language designs, so that new languages could support generic programming without pitfalls identified by Garcia et al.

Interpretations of generic programming vary depending on what kind of parameterization a programming language supports. Gibbons gives a taxonomy for some interpretations of genericity [31] which we reuse here. Programs parameterized by type constructors give rise to *genericity by shape*, through *datatype-generic programming* (also called *polytypism*) [1, 31]. This is the interpretation chosen by, e.g., Generic Haskell [52]. In the object-oriented world, generic programming refers primarily to *generics* or *parametric polymorphism* [11, 56], that is *genericity by type*. We add qualifiers such as *bounded* or *constrained* to these terms, and mean roughly the same things. Algebraic specifications are the basis for another approach to generic programming called *parameterized programming*. Parameterized programming has been concretized prominently in the OBJ family of languages, e.g. in OBJ2 [28], OBJ3 [37], CafeOBJ [24], and Maude [19, 20]. C++ *concepts* (as proposed for C++11), which describe syntactic and semantic requirements on data structures and algorithms [42], also descend from this approach based on algebraic specifications. Concepts, as implemented in C++20, only support syntactic requirements: we talk about *genericity by structure*. In the fully-fledged version of concepts, when both syntactic and semantic requirements are supported, we talk about *genericity by property*. C++ concepts were born out of Stepanov's work on generic programming [23, 60]. This paper, following Garcia et al., takes the notion of generic programming as introduced by Musser and

4:2

> **Generic programming** is a sub-discipline of computer science that deals with finding abstract representations of efficient algorithms, data structures, and other software concepts, and with their systematic organization. The goal of generic programming is to express algorithms and data structures in a broadly adaptable, interoperable form that allows their direct use in software construction. Key ideas include:
> - Expressing algorithms with minimal assumptions about data abstractions, and vice versa, thus making them as interoperable as possible.
> - Lifting of a concrete algorithm to as general a level as possible without losing efficiency; i.e., the most abstract form such that when specialized to the concrete case, the result is just as efficient as the original algorithm.
> - When the result of lifting is not general enough to cover all uses of an algorithm, additionally providing a more general form, but ensuring that the most efficient specialized form is automatically chosen when applicable.
> - Providing more than one generic algorithm for the same purpose and at the same level of abstraction, when none dominates the others in efficiency for all inputs. This introduces the necessity to provide sufficiently precise characterizations of the domain for which each algorithm is the most efficient.

■ **Figure 1** Definition of generic programming from Jazayeri, Musser, and Loos [58].

Stepanov in their seminal work in 1988 [69]. Figure 1 reproduces their structured definition of generic programming, taken from Jazayeri, Musser, and Loos [58].

We employ Garcia et al.'s framework for evaluating languages for generic programming to assess the approach based on algebraic specifications through an experiment with the Magnolia programming language [2]. This research language was first developed more than a decade ago, and is now again under active development [16]. We repeat Garcia et al.'s experiment of implementing a subset of the Boost Graph Library [82] (BGL), rich in generic definitions, to put the generic programming facilities of a language under rigorous scrutiny.

Magnolia is designed as an embodiment of a language for Stepanov-style generic programming. Magnolia's main type of genericity is thus *genericity by property*, as the language allows the specifications of algebraic signatures along with semantic requirements on their behavior, i.e., concepts. Magnolia does not offer any primitive type (beyond predicates), and it is designed to be parameterized by a host programming language and data structures implemented in that language. In the style of Gibbons's taxonomy, we coin the term *genericity by host language* to refer to the type of generic programming enabled by this axis of parameterization. One can implement composite operations in Magnolia—all base types and their operations, even loop structures, come from libraries written in the host language. Magnolia has a transpiler architecture, where the boundary between the language (Magnolia) and the base library (written in the host language) is not predefined, but rather the programmer can freely place it where convenient.

Garcia et al.'s work [30] to implement the same generic library in a variety of languages led the authors to identify several language properties that are useful and/or necessary for effective generic programming. Siek and Lumsdaine, in the context of developing the 𝒢 programming language, extended this set of properties [86]. These sets of properties served as the language evaluation framework in the above two works. We adopt this framework on the one hand to assess Magnolia's support for generic programming and on the other hand to relate its somewhat unorthodox language design to (more) mainstream languages. The listing of the identified properties, with our additions, is shown in Figure 2.

Following the recipe of the prior works, we implement a fragment of a generic graph library modeled after the BGL, in Magnolia, and analyze the result with regards to each identified property. This experiment allows us to extract several insights into generic programming, which we discuss in the paper. We highlight two particularly noteworthy aspects of the Magnolia BGL fragment. First, we implement both C++ and Python backend libraries. The same generic Magnolia code that captures the essence of graph algorithms can then be transpiled to either of these languages. We achieve an additional level of genericity, i.e., genericity by host language. Second, we show how a (seemingly) sequential generic algorithm can be transformed into one that is parallel, by picking appropriate backend data structures. This is achieved by abstracting the iteration mechanism. Magnolia does not offer any built-in looping constructs, and repetitions are thus necessarily expressed as generic abstractions.

The paper is structured as follows. Section 2 describes the landscape of languages designed for generic programming based on algebraic specifications, and explains how the approach is concretized, first in Maude and then in Magnolia. Section 3 presents our small graph library and discusses its Magnolia implementation. Section 4 situates Magnolia within the landscape of generic languages. It also makes connections and comparisons with other languages, and discusses related work. Section 5 discusses the performance of our implementation. Section 6 reflects on our approach and the insights we gained by developing the graph library. All the code discussed in this paper is made available [16].

## 2 Languages Designed for Generic Programming: The Approach of Algebraic Specifications

Algebraic specifications are at the core of Stepanov's work on generic programming [23, 60, 69, 87]. Highly influential early work in the field is Goguen's parameterized programming that emphasizes code reuse and modularity [29, 33]. Siek characterizes parameterized programming as similar to Stepanov's notion of generic programming, but without the same emphasis on efficiency [84]. Parameterized programming thus also aims at expressing algorithms in their most general form, making both their syntactic and semantic requirements explicit, and well organized.

### 2.1 Algebraic Specifications and Maude

Algebraic specifications and Goguen and Burstall's theory of institutions [36] have guided the design of the OBJ language family [39] (OBJ2, OBJ3, CafeOBJ, Maude...).

These languages provide extensive support for parameterized programming by design. OBJ2 and OBJ3 are both implementations of the OBJ logical programming language that differ in their operational semantics [37]. Maude incorporates most features of OBJ3 and significantly expands the capabilities of OBJ2 and OBJ3 for parameterized programming. Maude and CafeOBJ are still under active development. We describe below the general design of languages intended to support generic programming using algebraic specifications, and explain how it is concretized in Maude. Maude is based on rewriting logic [21, 34], and uses *membership equational logic* as its underlying equational logic. Our discussion only touches upon the fragment of Maude related to membership equational logic, where Maude's support for parameterized programming is concretized.

The general approach relies on a bilevel module system, with modules that allow for specifying generic APIs on the one hand and modules that allow for writing concrete programs on the other hand [38]. Modules of the same kind may be composed, and program modules can be parameterized by specification modules. Specifications consist of an algebraic signature defining sorts and (total and partial) operations, along with semantic requirements on their behaviour called *axioms*. Satisfaction relations can be expressed which describe how a program (or a specification) satisfies the requirements of a given specification.

Specifications are given in Maude through *functional theories*—Goguen introduced the notion of types as theories [35]. Functional theories allow expressing semantic requirements using *equations* and *conditional equations*. In addition, Maude allows the specification of *subsorting relations* along with *membership axioms*. This approach allows flexible control of partiality and declaring relationships between types, e.g., natural numbers and integers. The choice of implementing partiality using subsorting has consequences on other language features. For example, it poses restrictions on overloading and thus also on the ability to compose two arbitrary theories—see Listing 10 (in Appendix A) for an example. Note the similarity of this approach to *refinement types*—where the refined type $\{t : T \mid P\}$ is the subset of type $T$ for which the formula $P$ holds [27, 40]. Refinement types are closely related to subtyping.

Maude's *functional modules* allow for writing programs using the same constructs as functional theories—where equations and conditional equations define functions and data types in lieu of functional theories' semantic requirements, and where the rewriting system engendered by these equations must be confluent and terminating. The semantics of a functional module in Maude is the initial algebra defined by the module's equations, and evaluation is performed using an equational rewriting engine. Functional modules can be parameterized by functional theories: we speak of *parameterized functional modules*. Maude programs can be metarepresented as data and manipulated to produce new programs. This powerful mechanism of reflection allows generating so-called *dependent parameterized modules* such as $n$-tuples containing $n$ sorts and $n$ projection functions [19, Section 21.3.1]. Maude's built-in types are efficiently implemented in C++. Contrarily to the previous OBJ2 and OBJ3, Maude does not allow the user to implement custom primitive types in an external language.

Satisfaction relations in Maude are stated through *views*. Every sort (respectively function) in the view's source theory must be mapped (renamed) to a corresponding sort (respectively function) in the view's target module, and the mappings must preserve the subsorting structure of the source theory in the target module. It is also possible to implement functions on the fly to resolve signature mismatches.

## 2.2 Magnolia

As alluded to above, the Magnolia programming language is designed for Stepanov-style generic programming—i.e. parameterized programming with an added emphasis on efficiency. The language takes the same general approach based on algebraic specifications as described above, and its module system is likewise based on Goguen and Burstall's theory of institutions.

Listing 1 shows uses of the different module types. A **signature** allows defining types and operations. A **concept** is a **signature** augmented with **axiom**s that restrict the properties of the types and operations. A **concept** serves the same purpose as a functional theory in Maude, and the **signature** and **concept** modules constitute the specification layer of the module system. An **implementation** allows the same declarations as a **signature**, but also the definition of generic operation implementations; it is the equivalent of a parameterized functional module in Maude. A **program** is a specific kind of **implementation** in which all the specified operations and types are matched with (non-generic) concrete implementations; either Magnolia code that has a concrete implementation or an implementation in the base library in the host language. The **implementation** and **program** modules constitute the program layer of the module system. Constructs analogous to Maude's metaprogramming facilities are under investigation for Magnolia through Syntactic Theory Functors (STFs) [50] but the Magnolia compiler supports only specific instances of STFs at the moment [17].

Types (sorts) in Magnolia are opaque identifiers. One cannot explicitly parameterize them, nor can one define relations such as subtyping relations between them. Operations can be **function**s, **procedure**s, or **predicate**s. Procedure calls are prefixed with the **call** keyword, while function calls follow the usual uncurried call syntax. Predicates are treated as functions with a built-in, non-reimplementable return type. Magnolia's approach to partiality is based on guarded algebras [51]: an operation can be guarded by a predicate, which then acts as a precondition. In addition to their types, a **procedure** associates modes to its arguments: **obs** (read-only), **upd** (can be read and written to), and **out** (write-only, and must be written to) [5]. `ExampleProgram` in Listing 1 shows equivalent implementations of a multiplication by three as a **procedure** (`timesThreeUpdateRef`) and as a **function** (`timesThree`). In the example's program, the `int` type and `add` function are externally defined in Python and come from `PyConcreteSemigroup`. The line **use** `Magma[ T => int, bop => add ]` applies a renaming function to the content of the `Magma` **signature** and brings it into scope. The renaming maps `T` to a new name `int`, and `bop` to a new name `add`. It is assumed that the primitives implemented in the host language do not have side-effects, except for the modification of arguments to procedures where the argument mode is **out** or **upd**.

▪ **Listing 1** The main Magnolia building blocks.

```
1  signature Magma = {
2    type T;
3    function bop(t1: T, t2: T): T;
4  }
5
6  concept Semigroup = {
7    use Magma;
8    axiom bopIsAssociative(t1: T, t2: T, t3: T) {
9      assert bop(t1, bop(t2, t3)) == bop(bop(t1, t2), t3);
10   }
11 }
12
13 implementation PyConcreteSemigroup =
14   external Python lib.int_impl {
15     use Magma[ T => int, bop => add ];
16     use Magma[ T => int, bop => mul ];
17   }
18
19 program ExampleProgram = {
20   use PyConcreteSemigroup;
21   procedure timesThreeUpdateRef(upd i: int) {
22     i = add(add(i, i), i);
23   }
24
25   function timesThree(i: int): int {
26     var mutable_i = i;
27     call timesThreeUpdateRef(mutable_i);
28     value mutable_i;
29   }
30 }
31
32 satisfaction ExampleProgramHasAddSemigroup =
33   ExampleProgram models Semigroup[ T => int, bop => add ];
34
35 satisfaction ExampleProgramHasMulSemigroup =
36   ExampleProgram models Semigroup[ T => int, bop => mul ];
```

A **satisfaction** allows defining a modeling relation between an **implementation** and a **concept**; or between two **concept**s—it is the equivalent of a view in Maude. Signature mismatches are resolved through the renaming mechanism.

Magnolia semantics are tightly coupled to abstracting over hardware features: primitive types and operations may directly represent characteristics of the underlying hardware architecture, such as instruction sets, memory layout, etc. This enables Magnolia code to run efficiently on a variety of hardware, and to explore software for high-performance computing (HPC) [17]—making it suitable to address also the efficiency aspect of generic programming. This feature enables the user to utilize features of new hardware, e.g., posit numbers [44] by writing code directly in the targeted host language.

The notion of concepts, around which specifications in Magnolia are constructed, is from Stepanov and Musser [69]. These foundational building blocks of generic specifications and programs manifested in C++ first as mere documentation, then as library "hacks" [83, 89], and later as a language feature. The first proposals, see Siek's account of the history [85], were quite ambitious, including, e.g., semantic constraints (like Magnolia's axioms) and concept-based overloading, but their current form is somewhat scaled back. It is clear that concepts as a notion and language feature has been a highly influential contribution.[1]

We use Magnolia as a representative for languages designed for generic programming based on algebraic specifications throughout the remainder of the paper. The design of Magnolia and languages in the OBJ family draw from the same foundations, and the conclusions we draw about Magnolia should apply to these languages as well.

## 3  Graph Library in Magnolia

The subset of the BGL we implemented is a bit larger than the subset that Garcia et al. used. It consists of the six generic algorithms implemented by Garcia et al., i.e. Graph Search, Breadth-First Search (BFS), Dijkstra's single-source shortest paths, Bellman-Ford's single-source shortest paths, Johnson's all-pairs shortest paths, and Prim's minimum spanning tree, as well as a seventh algorithm, namely Depth-First Search (DFS). Like in Garcia et al.'s first study, we omit discussion of most algorithms for the sake of brevity—and discuss mainly our implementation of the BFS algorithm. This implementation is at the core of the library we implemented, and follows the same pattern as BGL's sequential implementation of BFS, whose pseudo-code is given in Listing 2.

---

[1] As a case in point, the 2021 ACM SIGPLAN International Conference on Software Language Engineering's "Most Influential Paper Award" was given to *Design of Concept Libraries for C++* by Sutton and Stroustrup [90], https://twitter.com/bcombemale/status/1449743946626221440.

We later show how careful choices in the instantiation of backend data structures allow using the same (seemingly) sequential BFS code to realize a parallel breadth-first graph traversal algorithm.

■ **Listing 2** Pseudo-code for the BFS algorithm implemented in the BGL [82]. Taken from https://www.boost.org/doc/libs/1_79_0/libs/graph/doc/breadth_first_search.html with minor stylistic changes.

```
BFS(G, s)
  for each vertex u in V[G] // initialize vertex u
    color[u] := WHITE
    d[u] := infinity
    p[u] := u
  end for
  color[s] := GRAY
  d[s] := 0
  ENQUEUE(Q, s) // discover vertex s
  while (Q != ∅)
    u := DEQUEUE(Q) // examine vertex u
    for each vertex v in Adj[u] // examine edge (u,v)
      if (color[v] = WHITE) // tree edge (u,v)
        color[v] := GRAY
        d[v] := d[u] + 1
        p[v] := u
        ENQUEUE(Q, v) // discover vertex v
      else // non-tree edge (u,v)
        if (color[v] = GRAY)
          ... // gray target (u,v)
        else
          ... // black target (u,v)
    end for
    color[u] := BLACK // finish vertex u
  end while
  return (d, p)
```

### 3.1 Implementing the Graph Algorithms

The BGL's implementation is based on the textbook BFS algorithm from Cormen et al.'s "Introduction to Algorithms" [22] that maintains the state of the traversal using a color map indexed by vertices. BGL's version adds to the algorithm various user-parameterizable visitor events, shown by the commented out actions in Listing 2. E.g., the "discover vertex" action is performed every time a vertex is encountered for the first time. The visitor events may modify the vertex queue (worklist) as well as an arbitrary user-provided state. By carrying around the right state and providing appropriate actions for each event, many algorithms can be built on top of the generic BFS implementation. Dijkstra's algorithm, for instance, can be implemented by carrying a state containing edge costs and vertex costs, and by updating the vertex costs every time an edge is examined.

The corresponding Magnolia implementation is split up into several functions across several modules and is rather lengthy. To improve readability, we put the full listings accompanying this section in Appendix A, and intersperse only excerpts with our text here. Listing 11 presents the `GenericBFSUtils` module, corresponding to lines 7 to 26 in Listing 2.

```
1 procedure breadthFirstVisit(obs g: Graph,
2     obs s: VertexDescriptor, upd a: A, upd q: Queue,
3     upd c: ColorPropertyMap) {
4   call discoverVertex(s, g, q, a);
5   call push(s, q);
6   call put(c, s, gray());
7   call bfsOuterLoopRepeat(a, q, c, g);
8 }
```

The entry point in `GenericBFSUtils` is `breadthFirstVisit`, which discovers the initial vertex and adds it to the queue, before calling `bfsOuterLoopRepeat`. The `bfsOuterLoopRepeat` procedure corresponds to the outer while loop in Listing 2 (lines 10 to 25), with the body of the loop implemented in `bfsOuterLoopStep` (reproduced below); we discuss this in more detail in Subsection 3.3.

```
1 procedure bfsOuterLoopStep(upd x: A, upd q: Queue,
2     upd c: ColorPropertyMap, obs g: Graph) {
3   var u = front(q);
4   call pop(q);
5   call examineVertex(u, g, q, x);
6   var edgeItr: OutEdgeIterator;
7   call outEdges(u, g, edgeItr);
8   call bfsInnerLoopRepeat(edgeItr, x, q, c, g, u);
9   call put(c, u, black());
10  call finishVertex(u, g, q, x);
11 }
```

Next is `bfsInnerLoopRepeat`, which corresponds to the inner for-each loop in Listing 2 (lines 12 to 23). The inner loop's body is implemented in `bfsInnerLoopStep` (see Appendix A).

The initialization of the queue and the color map is done in `search`, which is part of the `GraphSearch` module presented in Listing 3.

The `search` function is an entry point for simple graph searches in which an empty constructor that takes no argument exists for the queue. This is the case for a FIFO queue for instance, but not necessarily for a priority queue. For example, to implement Dijkstra's algorithm, we might want to use a priority queue that stores the shortest measured distance from the source to each vertex. The empty constructor for such a queue would take this information as a parameter—thus exposing a different API.

Listing 4 completes the implementation of the BFS: the types and operations are renamed and the underlying queue data structure is set to be a FIFO queue.

4:104:10

■ **Listing 3** Implementation of a graph search entry point in Magnolia.

```
implementation GraphSearch = {
  use GenericBFSUtils;
  require function empty(): Queue;

  function search(g: Graph, start: VertexDescriptor,
      init: A): A = {
    var q = empty(): Queue;
    var vertexItr: VertexIterator;
    call vertices(g, vertexItr);
    var c = initMap(vertexItr, white());
    var a = init;

    call breadthFirstVisit(g, start, a, q, c);
    value a;
  }
}
```

■ **Listing 4** Implementation of a BFS in Magnolia.

```
implementation BFS = {
  use GraphSearch[ search => breadthFirstSearch,
    Queue => FIFOQueue ];
  use FIFOQueue[ A => VertexDescriptor,
    isEmpty => isEmptyQueue ];
}
```

By keeping the requirements on the queue implementation loose in the `GraphSearch` module, we can produce a DFS implementation following the same pattern as in Listing 4—but using a LIFO queue (i.e., a stack) instead of a FIFO queue, and with appropriate renamings. Listing 12 (in Appendix A) shows how.

Dijkstra's algorithm is also implemented reusing the code in `GenericBFSUtils`, this time using a priority queue.

### 3.2 Specifying and Instantiating Data Structures

Both the FIFO queue and the stack concepts are easily derived from the generic `Queue` concept in Listing 5; the stack case is shown in Listing 6. Note that the concept of a stack exposes an operation named `top` instead of one named `front`. Thanks to the use of Magnolia's powerful renaming mechanism, this is not a problem: we can instantiate generic algorithms with data structures that provide the expected API up to renaming.

The concept in Listing 6 describes a stack by virtue of the axioms that refine a generic queue concept's behavior. Magnolia allows specifying axioms as part of concepts. They place restrictions on the behavior of operations' implementations. The `pushPopTopBehavior` axiom, for example, tells us that whenever a value a is pushed to any stack s, calling `top` on the resulting stack s′ yields a; similarly, calling `pop` on s′ yields s.

■ **Listing 5** Specification of a generic queue in Magnolia.

```
concept Queue = {
    require type A;
    type Queue;

    predicate isEmpty(q: Queue);
    procedure push(obs a: A, upd q: Queue);
    procedure pop(upd q: Queue) guard !isEmpty(q);
    function front(q: Queue): A guard !isEmpty(q);
}
```

■ **Listing 6** Specification of a generic stack in Magnolia.

```
concept Stack = {
  use Queue[ Queue => Stack, front => top ];

  function empty(): Stack;
  axiom pushPopTopBehavior(s: Stack, a: A) {
    var mut_s = s;
    call push(a, mut_s);
    assert top(mut_s) == a;

    call pop(mut_s);
    assert mut_s == s;
  }
  axiom emptyIsEmpty() {
    assert isEmpty(empty());
  }
}
```

Possible (hand-coded) user-provided backend data structure implementations for the stack concept of Listing 6 are given in Appendix A in Listings 13 (for C++) and 14 (for Python).

### 3.3 Abstracting the Schedule of the Algorithms

When comparing the Magnolia implementation to the pseudo-code in Listing 2, one can notice that the former has no loop structure. The outer (while) loop in the pseudo-code is implemented by a triplet of operations: `bfsOuterLoopCond`, which corresponds to the condition of the loop, `bfsOuterLoopStep`, which corresponds to the body of the loop, and `bfsOuterLoopRepeat`, which is called to start the loop. The inner for-each loop is implemented by a pair of operations, `bfsInnerLoopRepeat` and `bfsInnerLoopStep`.

Though this may seem tedious, it is by design that Magnolia provides no loop structure. The ideal manner to schedule and allocate computations (in a loop or otherwise) depends heavily on the hardware architecture, and by not having loops Magnolia forces this choice to remain a parameter, defined in a base library in the host language.

■ **Listing 7** Specification of a generic while loop in Magnolia.

```
concept WhileLoop = {
    require type Context;
    require type State;

    require predicate cond(s: State, c: Context);
    require procedure step(upd s: State, obs c: Context);
    procedure repeat(upd s: State, obs c: Context);

    axiom whileLoopBehavior(s: State, c: Context) {
        if cond(s, c) then {
            // if the condition holds, then doing one step and
            // completing the loop is the same as just completing
            // the loop
            var mutableState1 = s;
            var mutableState2 = s;
            call repeat(mutableState1, c);
            call step(mutableState2, c);
            call repeat(mutableState2, c)
            assert mutableState1 == mutableState2;
        }
        else {
            // otherwise, the state shouldn't change
            var mutableState1 = s;
            call repeat(mutableState1, c);
            assert mutableState1 == s;
        };
    }
};
```

A generic specification of a while loop in Magnolia is presented in Listing 7, and a corresponding C++ backend data structure implementation is shown in Listing 15—the latter listing can be found in Appendix A. The `WhileLoop` concept describes an API that takes two types (`Context` and `State`) and two operations (`cond` and `step`), and provides a `repeat` procedure whose behavior must correspond to the constraints expressed in the `whileLoopBehavior` axiom. By implementing projections on the opaque `Context` and `State` types, and updates on `State`, we can carry around arbitrarily complex contexts and states.

For the experiments described in Section 5, we implemented the loops of Listing 11 differently, to carry several state and context arguments. This was done for performance reasons, to avoid the overhead of packing and unpacking the `State` and `Context` objects. Magnolia lacks *variadics*, definitions that are generic on arity, i.e. on the number of arguments. Such a feature could let us avoid the packing and unpacking without the need to specify different concepts. We discuss this further in Section 4.

Abstracting away the loop structure (instead of providing a native Magnolia construct) has advantages: repetition can be implemented differently for different data structures or different target architectures. In our small BGL fragment, we exploited this aspect of Magnolia to provide two different backend implementations (in C++) for the inner for-each loop of the BFS algorithm: one that uses a sequential for loop and another that uses a parallel for loop based on OpenMP [73]. This is possible because the algorithm does not enforce a processing order on not-yet-visited vertices adjacent to the current vertex—the iterations of the inner loop are independent. The parallel version of the code also requires using explicitly thread-safe data structures for the vertex queue (and for the user-provided state, depending on how it is modified by the visitor events).

The difference in code when going from sequential to parallel is minimal: when concretizing our generic BFS algorithm, it suffices to **use** three modules exposing the same API as their sequential counterpart, but with different properties. By not committing to a looping mechanism too early, we gain a new powerful axis of parameterization.

## 4  Generic Features: Evaluation

With an understanding of the Magnolia implementation of the generic graph library, we can relate the code to the important language properties for generic programming identified by Garcia et al. [30] and Siek and Lumsdaine [86]. Figure 2 summarizes how Magnolia fares.

We should be cognizant that the list of properties is a reflection of the desire to express generic programs well in mainstream multi-paradigm languages, and maybe even based on experiences and programming idioms of C++. This is understandable: while Stepanov's and Musser's generic programming notions evolved through many languages, including Scheme and Ada, they materialized most prominently in C++. It is thus the case that even though the evaluation with the listed properties revealed shortcomings in programming languages, the properties arose from a C++-centric view of generic programming. Some are artifacts of this view and others more of a means to an end, rather than an essential part of a foundation for generic programming. Indeed, despite the several empty bullets in Magnolia's column in Figure 2, the BGL experiment was successful, likely because Magnolia builds its generics on somewhat different foundations than any of the languages studied by Garcia et al. (we like to think that it is closer to Stepanov's and Musser's ideals).

We also note that while the list of properties is rather comprehensive, we did end up adding two new items: variadics and property-based specifications. These are not relevant only to Magnolia, but would have been interesting topics of study in the original evaluation as well: variadic templates were studied after Garcia et al.'s evaluation [41] and the feature is today part of standard C++; property-based specifications (axioms) were proposed to be included in C++, e.g., for enabling optimizations, and the Haskell GHC compiler supports such specifications (in compiler pragmas) for rewriting [74].

For each of the properties listed in Figure 2, we give below its definition as given by Siek and Lumsdaine [86], motivate it briefly, and discuss its relevance and realization in Magnolia. We do not do any reimplementation for the previously studied languages. However, for the new properties we introduced, we also discuss their realization in the most recent release of the previously studied languages at the time of writing.

|  | C++ | SML | OCaml | Haskell | Java | C# | Cecil | C++0x | 𝒢 | Magnolia |
|---:|:---:|:---:|:---:|:---:|:---:|:---:|:---:|:---:|:---:|:---:|
| Multi-type concepts | - | ● | ○ | ● | ○ | ○ | ◐ | ● | ● | ● |
| Multiple constraints | - | ◐ | ◐ | ● | ● | ● | ● | ● | ● | ● |
| Associated type access | ● | ● | ◐ | ● | ◐ | ◐ | ◐ | ● | ● | ● |
| Constraints on associated types | - | ● | ● | ● | ◐ | ◐ | ● | ● | ● | ● |
| Retroactive modeling | - | ● | ● | ● | ○ | ○ | ● | ● | ● | ● |
| Type aliases | ● | ● | ● | ● | ● | ○ | ○ | ● | ● | ● |
| Separate compilation | ○ | ● | ● | ● | ● | ● | ◐ | ○ | ● | ○ |
| Implicit argument deduction | ● | ○ | ● | ● | ● | ● | ◐ | ● | ● | ○ |
| Modular type checking | ○ | ● | ◐ | ● | ● | ● | ◐ | ◐ | ● | ● |
| Lexically scoped models | ○ | ● | ○ | ○ | ○ | ○ | ○ | ● | ● | ● |
| Concept-based overloading | ● | ○ | ○ | ○ | ○ | ○ | ● | ● | ◐ | ○ |
| Same-type constraints | - | ● | ○ | ● | ○ | ○ | ○ | ● | ● | ○ |
| First-class functions | ○ | ● | ● | ● | ○ | ◐ | ● | ● | ◐ | ○ |
| Property-based specifications | ○ | ○ | ○ | ◐ | ○ | ○ | ○ | ● | ○ | ● |
| Variadics | ● | ○ | ● | ● | ◐ | ◐ | ○ | ● | ○ | ○ |

■ **Figure 2** The level of support in Magnolia for properties for generic programming. For the reader's convenience, we reproduce here the original characterization of C++, SML, OCaml, Haskell, Java, C#, Cecil, C++0x, and 𝒢 from Siek and Lumsdaine [86] (omitting footnotes with detailed commentary). ● indicates full support, ○ indicates poor support, and ◐ indicates partial support. The rating of "-" for C++ indicates that while C++ does not explicitly support the feature, one can still program as if the feature were supported. The level of support for *property-based specifications* and *variadics* is indicated for the latest release of each language at the time of writing, i.e. respectively, for the first seven columns, C++20, SML'97, OCaml 4.14, Haskell 2010, Java 18, C# 11, and Cecil 3.2 [15]. We evaluate C++0x as it was envisioned, as opposed to its eventual partial adoption in C++11. To the best of the authors' knowledge, 𝒢 has only had one release.

**Multi-type concepts**
*A concept can be implemented by a collaboration of several types.*

Multi-type concepts in generic programming correspond to multi-sorted signatures with axioms in algebraic specifications, and both arise naturally and often. Magnolia's concepts can declare any number of types and define syntactic and semantic requirements on any combination of them. Further, the partial order of concepts that arises from Magnolia concept definitions and their **use** declarations is not in any way constrained by the types declared in the concept that *uses* or the concept that is being *used*. Any name conflicts that might arise are easily resolved with renaming types and operations. Magnolia thus fully supports multi-type concepts.

By contrast, in many other languages, in particular in object-oriented ones, concepts are approximated by interfaces/classes, which are types. These interface/class types are treated differently from other types of the concept (defined as type parameters to the generic interface), which introduces many restrictions, and obstacles for the clean expression of generic programs [30, 57].

**Multiple constraints**
*More than one constraint can be placed on a type parameter.*

A few of the languages studied by Garcia et al. had restrictions when constraining types by more than one concept; the reasons are technical, and discussed in prior work [30]. In Magnolia there are no restrictions: *multiple constraints* merely mean that a particular type appears in more than one **use** declaration. All definitions from used concepts are brought to the same scope; the type's constraints are thus a union of its requirements in these concepts.

**Associated type access**
*Types can be mapped to other types within the context of a generic function.*

In most object-oriented languages studied by Garcia et al., the only way to declare types of a concept is as type parameters of a generic. The evaluation called for a mechanism for defining type members, "associated" types of the "main" type(s) of the concept. One could do this in C++ with trait classes and in Haskell, at the time with the *functional dependencies* extension, later with associated types [14]. Today, Rust and Swift also have similar notions. Associated types solve problems of instantiating concepts with positional type parameters, e.g., that they can shorten the parameter lists considerably. Järvi et al. [57] detail these problems in Java and C#.

In Magnolia, there is no distinction between main types and associated types. All types are opaque, and accessible by their name. Magnolia **use**s of concepts only mention the types (and operations) that the programmer needs or wants to rename—there is no need to anticipate which types are better expressed as main types, which as associated types.

**Constraints on associated types**
*Concepts may include constraints on associated types.*

Declaring constraints on associated types leads to problems in several languages; Java and C#, for example, require redundant constraints (for complex technical reasons [57]). As discussed above, all types in a concept are treated the same in Magnolia, and hence Magnolia supports constraints on associated types as it does for any type.

**Retroactive modeling**
*New modeling relationships can be added after a type has been defined.*

Problematic languages concerning this property are languages where the declaration that a data type satisfies a certain set of requirements (a concept or concepts) takes place at the site of definition of the data type. This is the case for object-oriented languages that fix the bases of a class when the class is defined. But even in Haskell, where an *instance declaration* is distinct from both a datatype and a type class definition, retroactive modeling can be limited. An example is changes to Haskell's standard library and its type class hierarchy. In 2014 the `Applicative` typeclass was suggested to be made a superclass of the `Monad` typeclass [46]. Such a change breaks `Monad` instances (models) where the corresponding `Functor` and `Applicative` instances are not implemented. This change occurred after the study of Garcia et al. [30], and thus likely was not considered when evaluating the support of Haskell for retroactive modeling—the study characterized Haskell as fully supporting retroactive modeling.

Listing 8 builds up to the concepts of a commutative magma with a left absorbing element, and of a commutative magma with a right absorbing element. The two concepts are equivalent, i.e., each concept models the other. Each modeling relationship is expressed through a satisfaction relation (see `CommutativeZeroLR` and `CommutativeZeroRL`). Satisfaction relations can be added at any point in the program, hence Magnolia satisfies the retroactive modeling property.

**Type aliases**
*A mechanism for creating shorter names for types is provided.*

The problem of long names in generic programming often arise from a large number of type parameters (due to the representation of associated types as type parameters). In Magnolia, concepts are not represented by types—type names are not parameterized, their names thus stay atomic. Magnolia does support type aliases too, however, through the mechanism of renaming — see once again lines 15 and 16 in Listing 1. Magnolia does not allow declaring a new alias for a type (or operation) in the same module expression. That being said, it is possible to make declarations with different names in a module expression, and to later merge them using renaming, therefore "retroactively" aliasing the two declarations.

**Separate compilation**
*Generic functions can be compiled independently of calls to them.*

The motivation behind separate compilation is attaining better compilation speed by avoiding recompiling generic definitions every time their uses are compiled. All languages but C++ in Garcia et al.'s evaluation have this compilation model; in C++ the compiler generates a distinct piece of code for each different template instantiation. While Garcia et al. bundled separate compilation and modular type checking under one language property, Siek and Lumsdaine [86] split them into two distinct properties.

**Revisiting Language Support for Generic Programming**■ **Listing 8** An example of equivalent specifications in Magnolia.

```
concept CommutativeMagma = {
  type T;
  function bop(t1: T, t2: T): T;
  axiom commutativity(t1: T, t2: T)) {
    assert bop(t1, t2) == bop(t2, t1);
  }
}

concept CommutativeMagmaWithLeftZero = {
  use CommutativeMagma;
  function zero(): T;
  axiom leftAbsorption(t: T) {
    assert bop(zero(), t) == zero();
  }
}

concept CommutativeMagmaWithRightZero = {
  use CommutativeMagma;
  function zero(): T;
  axiom rightAbsorption(t: T) {
    assert bop(t, zero()) == zero();
  }
}

satisfaction CommutativeZeroLR =
  CommutativeMagmaWithLeftZero models
    CommutativeMagmaWithRightZero;

satisfaction CommutativeZeroRL =
  CommutativeMagmaWithRightZero models
    CommutativeMagmaWithLeftZero;
```

This allows to more precisely characterize C++ after the concepts feature was added— C++ today partially supports modular type checking but not separate compilation.

In Magnolia, generic operations are type checked where they are declared. They may undergo name changes during renamings, but after these are resolved, a call to a generic function needs only to be checked against the function's declaration, so Magnolia supports modular type checking. Adhering strictly to the definition given above, Magnolia could be said to support separate compilation: each monomorphic operation is transpiled to the host language independently of calls to it (and the compilation of transpiled Magnolia code to executable code is host language-dependent). However, a distinct piece of code is emitted for each instantiation of a generic function definition — resulting in a compilation model similar to C++'s, and not one which achieves the goals of the property.

4:18

**Implicit argument deduction**
*The arguments for the type parameters of a generic function can be deduced and do not need to be provided by the programmer. Also, the finding of models to satisfy the constraints of a generic function is automated by the language implementation.*

Garcia et al. describe the lack of implicit type argument deduction to result in verbose generic algorithm invocations. Most languages avoid problems by deducing the type parameters of a generic function from the types of its function arguments. Magnolia avoids this problem in a different way: there are no implicit arguments to deduce in the first place. Operations are always monomorphic, and each argument's type is resolved whenever renaming occurs. Whenever a call occurs, there is almost always a single corresponding prototype in scope. The exception is when a call can resolve to several functions overloaded solely on their return type. In that case, a type annotation must be provided by the user to disambiguate between the matches.

**Modular type checking**
*Generic functions can be type checked independently of calls to them.*

Because of C++'s lack of modular type checking, debugging type errors in generic code in C++ is often very difficult. The C++ concepts language feature fixes this problem partially: uses of templates are checked against type parameter constraints but definitions of templates are not checked. The bodies of C++ template functions are (still) type checked only after their instantiation, which can delay catching a type errors in the implementation of a generic library until it is used in client code. Magnolia supports modular type checking of both the uses and definitions of generic code, as described above in the discussion of *Separate compilation*.

**Lexically scoped models**
*Model declarations are treated like any other declaration, and are in scope for the remainder of the enclosing namespace. Models may be explicitly imported from other namespaces.*

Siek and Lumsdaine implement lexically scoped models in $\mathscr{G}$ [86], in order to be explicit about which models are in scope. One motivation for this feature is to avoid the problem of overlapping models (corresponding to overlapping instances in Haskell, with concepts corresponding to typeclasses, models to typeclass instances). Suppose we want to define an **instance** of the `Semigroup` typeclass for `Int` in Haskell. The Haskell 2010 Language Report [65, Chapter 4] dictates that *"A type may not be declared as an instance of a particular class more than once in the program."*. However, there is more than one intuitive instance of `Semigroup` for `Int`, as shown in Listing 9.

Attempting to call (`<>`) with both of these definitions in scope results in an error. This can be worked around in awkward ways, e.g. using `newtypes`, or wrappers around class methods [94]. The crux of the issue here is that typeclass instances are not first-class in Haskell.

It is not clear if the *lexically scoped models* property is actually sufficient to solve the problem of overlapping models. The approach works well when the different models are used in different scopes, but does not seem to offer a solution when one wants to have them in the same scope. One example of such a use case is *united*

■ **Listing 9** Overlapping instances in Haskell.

```haskell
-- A.hs
instance Semigroup Int where
  (<>) = (+)
-- B.hs
instance Semigroup Int where
  (<>) = (*)
-- C.hs (imports A, B)
-- error: Overlapping instances for Semigroup Int
val = (2 :: Int) <> 3
```

*monoids*, an algebraic structure involving two monoids with the same unit element [68]. Approaches such as named instances [59] or the *CONCEPT* pattern in Scala [71] can address this issue.

The issue does not arise in Magnolia either: one can bring two different models of the same concept into the same scope and resolve the overlap explicitly using the renaming mechanism—see `ConcreteSemiGroup` in Listing 1 for an example. The property is thus supported by design.

**Concept-based overloading**
*There can be multiple generic functions with the same name but differing constraints. For a particular call, the most specific overload is chosen.*

The C++ standard library's hierarchy of iterator concepts includes two concepts, *InputIterator* and *ForwardIterator*, whose signatures agree—they differ only on their operations' semantic requirements. In particular, the former does not admit restarting the iteration. There are well-motivated cases for overloading a function where the overloads should be differentiated based only on whether their argument types model one or both of these concepts [88]. However, overloading on semantics is problematic in the general case. To specialize based on the semantics of two concepts with exactly the same API, the compiler needs to partially order them. Consider the two concepts `CommutativeMagmaWithRightZero` and `CommutativeMagmaWithLeftZero` from Listing 8. If one specializes an algorithm on their semantics, the following happens:

- if the compiler is unable to deduce that the specifications are equivalent, one specialization is picked at each call site and compilation succeeds;
- if the compiler is able to deduce that the specifications are equivalent, the compiler cannot specialize and compilation fails.

This implies that once correct code may become incorrect as the compiler's reasoning abilities get more powerful and it can deduce more properties from the same axioms. And indeed, C++ does not really do overloading on semantics. It equips each iterator concept with a *tag*-type, creating thus a syntactic difference between the two concepts' requirements, which is really what is used as a criterion in overload resolution.

Magnolia does not provide support for concept-based overloading. Arguments to an operation are always instances of the exact types specified in the operation's prototype. In the absence of subtyping, classic overloading is sufficient to resolve every call to the right implementation. The modular structure of Magnolia code allows the programmer to defer the implementation of types and operations as long as is necessary to sufficiently refine their semantic requirements and explicitly determine which implementation should be chosen.

**Same-type constraints**

The notion of *same-type constraints* lacks a precise definition in Siek and Lumsdaine's work. We give it the following definition: *It is possible to force two type parameters to refer to the same type.*

In Magnolia, two types are the same if they have the same name. The renaming mechanism allows forcing different concepts to depend on the same type, by bringing them into the same scope and renaming their type parameters to the same name. This is slightly different from a type constraint: instead of requiring two constrained type parameters, the resulting module has a single type parameter. This mechanism is demonstrated lines 15 and 16 in Listing 1: the `Magma` module is brought into scope twice, and its type parameter `T` is forced to the same name `int`.

**First-class functions**

*Supporting anonymous functions with lexical scoping as first-class citizens of the language.*

Magnolia does not support higher-order functions. This is intentional: it keeps Magnolia programs simpler to reason about. We do not lose out on expressivity—in lieu of higher-order functions, the Magnolia programmer can use a parameterized module [32], and deal with potential naming conflicts by leveraging once more the renaming mechanism. This is, however, certainly a trade-off on convenience. For example, the most cumbersome aspects of implementing the graph library were the looping structures, split into several concepts and functions (see e.g. Listings 7 and 15), which with higher-order functions could have been implemented with a single function parameterized by a function parameter.

**Property-based specifications**

We define the property as follows: *Arbitrary semantic constraints on types and operations can be defined.*

Property-based specifications are a desirable feature that can enable strong correctness guarantees and formal verification of code. Such semantic constraints are mentioned in Garcia et al.'s study [30], but they are not evaluated due to the lack of support for them in the studied languages. The way we specify properties in Magnolia axioms is through assertions—and in fact, any programming language with assertions can produce some sort of library support for property-based specifications. We explicitly do not consider this to qualify as language support for property-based specifications in Figure 2, but we mention some such libraries below.

**C++20** implements a scaled back version of C++0x's concepts which does not provide support for semantic constraints—but only for same-type constraints and API modeling constraints. Bagge et al. previously built a testing system atop concepts and axioms implemented using template metaprogramming in C++11 [4].

**SML** does not support property-based specifications. We note that property-based specifications for SML programs can be expressed in Sannella and Tarlecki's Extended ML [77]. However, the semantics of Standard ML are not fully compatible with the theory of algebraic specifications, and the approach suffers from a semantic gap common in many approaches to formal verification of software [78].

**OCaml** does not support property-based specifications. Xu showed how OCaml could be augmented with a contract declaration construct, along with both static and dynamic contract checking features [92]. However, to the best of the authors' knowledge, this research did not lead to the implementation of such a feature in OCaml. Design by contract (DbC) is a common approach to software correctness made popular by Eiffel [66, 67]. DbC has roots in Floyd-Hoare logic [26, 54] and uses assertions to specify *preconditions*, *postconditions*, and *invariants* on programs. Bagge et al. point out limitations with pre/postconditions for specifying generic APIs, e.g., difficulties of capturing properties like *associativity* or *transitivity*, and show how they are subsumed by axioms [6].

**Haskell**'s support for property-based specifications is limited. One visible consequence of this is that typeclass laws are typically stated only as documentation, and it is up to the programmer of a typeclass instance to ensure that they hold. However, the language's powerful type system and extensions allow specifying and enforcing sophisticated invariants. For example, Bailey and Gale encoded the full FIDE ruleset at the type level [7]. Noonan shows a design concept for validating preconditions at compile time by constructing proofs inhabiting phantom type parameters [70]. Haskell also offers good support for property-based testing, through the QuickCheck library [18]. Like for OCaml, some work on enabling static contract checking in Haskell was initiated, but did not lead to the implementation of such a feature in the language to the best of the authors' knowledge [93]. Also worthy of note is LiquidHaskell, a static verifier for Haskell based on liquid types [76]. Liquid types are refinement types [27] with logical predicates coming from a decidable sublanguage—allowing decidable type checking and inference.

**Java** itself does not support property-based specifications. However, there exist a number of tools extending Java to provide varying levels of support for property-based specifications. One such tool is the Java Modeling Language (JML), which draws from the design by contract approach and from algebraic specifications to allow for specifying the behavior of Java modules [64]. Another one is JAxT, a tool that generates JUnit test cases from static methods representing axioms [47, 49]. In addition, there exists several libraries implementing property-based testing à la QuickCheck for Java—one example is junit-quickcheck [55].

**C#** itself does not provide support for property-based specifications. Similarly to Java, several tools exist that address this shortcoming. For example, Spec# is a superset of C# which allows specifying and verifying method contracts (pre- and postconditions), object invariants, and loop invariants [8]. Code contracts are another approach that enables design-by-contract programming in .NET programming languages [25].

**Cecil** does not provide support for property-based specifications.

Axioms were supported in **C++0x** concepts, and the language had full support for property-based specifications.

Siek and Lumsdaine's 𝒢 [86] does not implement such semantic constraints, restricting itself to same-type constraints and API modeling constraints (similarly to the concepts implemented in C++20).

**Magnolia** supports property-based specifications in the form of axioms in concepts. These formulas are boolean expressions with free variables, thus encompassing equational and conditional equational specifications as common in algebraic specifications [10]. Boolean expressions have the benefit of being readily handled by programming language compilers and tools, allowing us to compile axioms as test oracles and systematically test a program's compliance with its specification [4]. The axiom formalism and the program code are semantically compatible, thus avoiding the semantic gap mentioned previously [78]. Magnolia axioms can be leveraged in practice for program optimizations [17] and for proving the correctness of Magnolia specifications [45].

**Variadics**

We define the property as follows: *Operations can have a variable number of arguments of different types.*

**C++ and C++0x** allow generic operations to take in a variable number of arguments of different types through variadic templates [41]. In a variadic context, any type expression can be repeated, including expressions containing the *const* qualifier, the lvalue and rvalue reference declarators, or any *concept* constraint. We note that variadic templates were introduced in C++11, a version of the language that postdates both Garcia et al.'s and Siek and Lumsdaine's studies. The version of C++ evaluated in the previous studies did not support variadics, but C++0x did.

It is possible to implement functions that support a variable number of arguments of different types in **OCaml**, as demonstrated in the current implementation of the *Format* module [91]. This implementation of variadics relies on heterogeneous lists, which are in turn implemented in OCaml with difference lists leveraging Generalized Algebraic Data Types (GADTs) [75, 79]. There are limitations to this approach, related to the mixing of GADTs and subtyping [75, 80].

Like for OCaml, there is no obvious way to define functions that take a variable number of arguments of different types in **Haskell** [53]. It is however possible to define variadic functions through type hackery, as demonstrated by the HList library [61, 63]. Haskell's expressivity allows for several reasonable ways to create such functions—as noted in the source code of HList [62]. Since 2015, heterogeneous lists are implemented using a *data family* [81].

Support for variadics in **Java** is only partial. It is possible to define operations that take a variable number of arguments of different, arbitrary types in Java by adding an argument of type `Object...` to the end of their argument list. The ellipsis syntax is syntactic sugar for passing in a single-dimensional array of the specified type as an argument. `Object` is a superclass for every defined class in Java, ensuring that the function can be called with parameters of any object type. Note that this excludes primitive types, which are not subclasses of `Object`, and for which there is no obvious solution.

Support for variadics in **C#** is limited. It is possible to define functions that take a variable number of arguments of different types in C# by using the `params` keyword to pass in an arbitrary number of arguments of type either `object` or `dynamic`. The `params` keyword is syntactic sugar for passing in an array as a parameter. Arguments given the `dynamic` type can not be type checked at compile time, and will cause run time exceptions if used inappropriately. Arguments converted to the `object` type must eventually be unboxed to the correct type to be used. In this case, some errors can be caught at compile time, and others at run time. It is also not possible to pass a variable number of generic type parameters to a function.

None of **SML**, **Cecil**, and 𝒢 [84, 86] (to the best of our knowledge) offer support for variadics.

Because all types are opaque in **Magnolia**, data structures are characterized only on the set of externally-implemented functions that construct or consume them. To define a record-like type with $n$ fields in Magnolia requires one type definition, along with $2n + 1$ function definitions (one projection and one update for each field, and a constructor). This quickly leads to a large number of functions. These projections and updates may also be expensive, as discussed in Subsection 3.3; there, we solved both problems by defining several loop concepts and backend implementations, each with a carefully chosen number of state and context types and parameters. This solution has its own drawbacks though: concepts and implementations are (mostly) duplicated, including axioms. Adding support for variadics to Magnolia would achieve the same outcomes, while eliminating the need for code duplication. We briefly discuss an approach for supporting variadics in Magnolia in Section 6.

## 5 Performance

Another key idea in generic programming is that abstracting an algorithm should have no impact on performance: when a generic algorithm is specialized to the concrete case, it should be just as efficient as if the algorithm had been written directly as the non-generic case. We tested whether Magnolia and its BFS implementation satisfy this criterion. Figure 3 compares the performance of two instantiations of our BFS implementations; for both, C++ was the host language. The figure also shows the performance of BGL's BFS implementation.

The two Magnolia implementations use the same generic algorithm with different backend data structures. The red bars show the performance of an instantiation that uses the same data structures as the BGL's algorithm (blue bars). The yellow bars show an instantiation of the Magnolia code that uses our own ad-hoc, prototype data structures. The transpiled algorithms are identical to the one implemented in the BGL. When using the same underlying data structures, the Magnolia and BGL C++ implementations perform equally well, showing that our generic abstraction in Magnolia is indeed cost-free.

Instantiating the algorithm with our prototype data structures produces code that runs roughly 2.5–3.5 times slower, depending on the number of edges in the graph. At the same time, these data structures offer more flexibility to the user by virtue of being more parameterizable—highlighting a trade-off between parameterization and performance here. Magnolia allows fine-grained choice of level of abstraction on the host language. This makes exploring the possible combinations of backend data structures easy. No particular effort was put in tuning our prototype data structures for performance: it is entirely possible that a more careful and as parameterizable design could match the BGL implementation's performance.

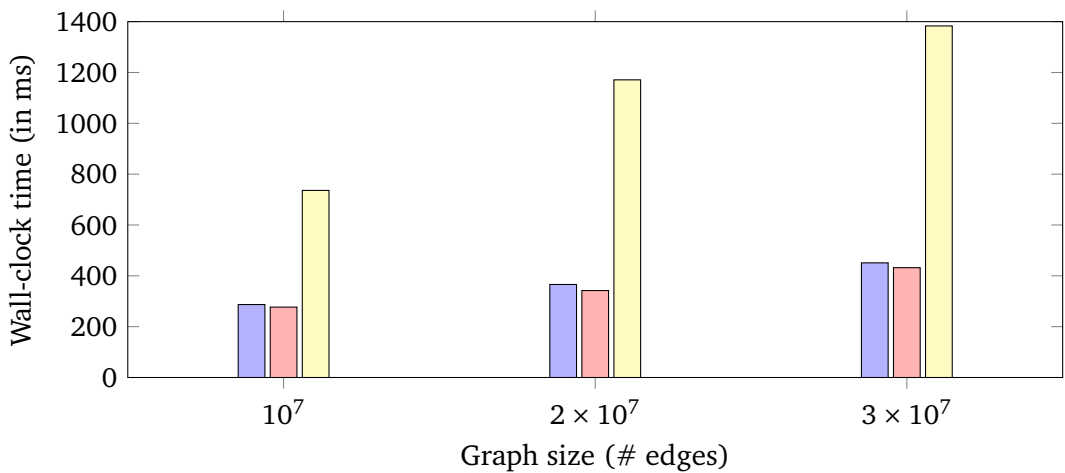

**Figure 3** Performance comparison of running the sequential version of the BFS algorithm: the leftmost columns (blue) are for BGL's C++ implementation, the middle (red) ones for the Magnolia implementation where the base library uses the same data structures as BGL, and the rightmost (yellow) ones for the Magnolia implementation that moves the language-library border further towards the language, providing highly parameterizable data structures. Both Magnolia implementations used C++ as the host language. Each implementation is run 10 times in total, and the running times are averaged. Every implementation is tested against the same 10 randomly generated graphs. Each graph is directed and contains $10^6$ vertices. The test programs are compiled using g++ 10.2.0, with optimization level O3, on an Intel(R) Xeon(R) Silver 4112 CPU @ 2.60GHz.

The transpilation of our Magnolia code to a host language does not add much overhead: it takes less than a second to transpile the whole fragment of the BGL we implemented. In contrast, compiling the final binary from the C++ code which imports the BGL takes more than seventeen seconds.[2]

We did not run performance tests with Python as the host language. We can expect the current implementation to be slow, because we left overload resolution to be performed in Python (out of convenience), using multiple dispatch.

---

[2] Compilation times reported for an Intel(R) Core(TM) i5-7300U CPU @ 2.60GHz.

## 6 Discussion and Conclusion

Garcia et al.'s study [30] paved the way for evaluating support for generic programming of languages. The properties the authors identify as important point out issues of retrofitting generic programming into existing languages. This in fact led language implementors to address these issues and improve their language's support for generic programming [13, 14]. Siek and Lumsdaine's 𝒢 demonstrates how a language based on these properties enables generic programming. Our work takes a step back from these works and looks at generic programming from the angle of algebraic specifications, repeating the experiment [30] with Magnolia—a language representative of the algebraic approach. Magnolia is not shoehorned into the properties that Garcia et al. identified, yet provides extensive support for generic programming.

Our evaluation in Section 4 shows that the renaming mechanism plays a crucial role in enabling generic programming in Magnolia. Renaming is Magnolia's pragmatic version of signature morphisms. It allows control over the naming of types and operations, both to keep them separate as needed for implementations, but also to coordinate naming within concepts when joining them together. This is somewhat less powerful than the signature morphisms supported by CASL [10], yet powerful enough to enable a high level of reuse between modules. Carette et al. recently investigated union and renaming as a reuse mechanism for modular specification of mathematical concepts [12].

Every programming language is its own formal system with its own advantages and inconvenients, and Magnolia is no exception. For example, while the algebraic approach gives extreme flexibility when it comes to parameterizing and combining modules, this flexibility comes at a usability cost: when developing in Magnolia, it is hard to keep track of what is in scope at a given line and where declarations come from. The problem is further exacerbated by the renaming mechanism. Tool support (e.g., in the form of an IDE) is crucial for Magnolia development. Bagge described an implementation of an IDE for Magnolia integrated with Eclipse [3]. The newer magnoliac compiler provides a basic interactive toplevel that allows users to inspect the content of loaded modules [16]. The design and development of a fully-fledged IDE for Magnolia will inform on whether the reasoning problems we faced when implementing the BGL in Magnolia can be mitigated, and is a topic of future work for us.

In Gibbons' taxonomy of generic programming [31] we characterized Magnolia as supporting genericity by property. This axis of genericity has been an inspiration of new features for C++ (concepts), and there has been expectations that proper language support for expressing semantic properties (axioms in concepts) will lead to domain-specific optimization opportunities, more precise static checking of code for semantic errors, and more flexible (concept-based) overloading. This experiment with Magnolia accentuates some challenges that will remain, even with full language support for properties. In particular, in our evaluation we discuss concept-based overloading and why overloading based on semantic properties is problematic. Further, there are challenges with the expression of semantic constraints in concepts. Sometimes axioms are not expressible by using solely the operations that a concept is meant to expose—additional operations need to be added to the concept just to be able to express a semantic property [9]. Listing 16 gives an example: given g: Graph, vertices(g) returns the collection of all vertices in g. Given v: Vertex a vertex of g, adjacentVertices(g, v) returns the collection of all the vertices adjacent to v in g. There is a subset relation between adjacentVertices(g, v) and vertices(g). To state this property through an axiom, we would need additional operations on VertexCollection, e.g., the ability to check whether a Vertex is a member of a VertexCollection. These operations and the axiom are shown as commented out.

Another challenge we identified with Magnolia is the inability to express variadic generic definitions. The general mechanism of syntactic theory functors [50] seems well-suited for implementing variadics in Magnolia. In fact, also renaming can be expressed as STFs. These connections are topics for future work for us.

**Acknowledgements** We offer our thanks to the anonymous reviewers for their thoughtful insights on our paper. We also thank Mikhail Barash for proofreading, and providing insightful comments about our manuscript in early stages.

## A  Code Listings

**Listing 10** Example of tension between overloading and subsorting in Maude specifications (adapted from an example by Ölveczky [72, Chapter 2.5]). When calling $f$ on an argument of type $s12$, Maude can not determine which overload of $f$ should be called.

```
fth OVERLOADING is
   sorts s1 s2 s12 u1 u2 .
   op f : s1 -> u1 .
   op f : s2 -> u2 .
endfth

fth SUBSORTING is
   including OVERLOADING .
   subsorts s12 < s1 s2 . --- error!
endfth
```

**Revisiting Language Support for Generic Programming**

■ **Listing 11** Implementation of generic BFS utils in Magnolia.

```
implementation GenericBFSUtils = {
  /* snip types and helper operation declarations */
  procedure breadthFirstVisit(obs g: Graph,
      obs s: VertexDescriptor, upd a: A, upd q: Queue,
      upd c: ColorPropertyMap) {
    call discoverVertex(s, g, q, a);
    call push(s, q);
    call put(c, s, gray());
    call bfsOuterLoopRepeat(a, q, c, g);
  }

  predicate bfsOuterLoopCond(a: A, q: Queue, c: ColorPropertyMap,
      g: Graph) { value !isEmptyQueue(q); }

  procedure bfsOuterLoopStep(upd x: A, upd q: Queue,
      upd c: ColorPropertyMap, obs g: Graph) {
    var u = front(q);
    call pop(q);
    call examineVertex(u, g, q, x);
    var edgeItr: EdgeIterator;
    call outEdges(u, g, edgeItr);
    call bfsInnerLoopRepeat(edgeItr, x, q, c, g, u);
    call put(c, u, black());
    call finishVertex(u, g, q, x);
  }

  procedure bfsInnerLoopStep(obs edgeItr: EdgeIterator,
      upd x: A, upd q: Queue, upd c: ColorPropertyMap,
      obs g: Graph, obs u: VertexDescriptor) {
    var e = edgeIterUnpack(edgeItr);
    var v = tgt(e, g);
    call examineEdge(e, g, q, x);
    var vc = get(c, v);
    if vc == white() then {
      call treeEdge(e, g, q, x);
      call put(c, v, gray());
      call discoverVertex(v, g, q, x);
      call push(v, q);
    } else if vc == gray() then {
      call grayTarget(e, g, q, x);
    } else { // vc == black();
      call blackTarget(e, g, q, x);
    };
  }
}
```

■ **Listing 12** Implementation of a DFS in Magnolia.

```
implementation DFS = {
  use GraphSearch[ search => depthFirstSearch, front => top,
    isEmptyQueue => isEmptyStack, Queue => Stack ]; // LIFOQueue
  use Stack[ A => VertexDescriptor, isEmpty => isEmptyStack ];
}
```

■ **Listing 13** User-provided (hand-coded) implementation of a stack in C++.

```cpp
template <typename _A>
struct stack {
  typedef _A A;
  typedef std::stack<A> Stack;

  Stack empty() { return Stack(); }
  bool isEmpty(const Stack &s) { return s.empty(); }
  void push(const A &a, Stack &s) { s.push(a); }
  void pop(Stack &s) { s.pop(); }
  const A &top(const Stack &s) { return s.top(); }
};
```

■ **Listing 14** User-provided (hand-coded) implementation of a stack in Python.

```python
def stack(A):
  class Stack:
    def __init__(self): self.stack = []
    def isEmpty(self): return not self.stack
    def push(self, a: A): self.stack.insert(0, a)
    def pop(self): self.stack = self.stack[1:]
    def top(self): return deepcopy(self.stack[0])
    def mutate(self, other): self.stack = other.stack[:]
  def empty(): return Stack()
  def isEmpty(s: Stack): return s.isEmpty()
  def push(a: A, s: Stack): s.push(a)
  def pop(s: Stack): s.pop()
  def top(s: Stack): return s.top()

  stack_tuple = namedtuple('stack',
    ['A', 'Stack', 'empty', 'isEmpty', 'push', 'pop', 'top'])

  return stack_tuple(A, Stack, empty, isEmpty, push, pop, top)
```

■ **Listing 15** User-provided (hand-coded) implementation of a while loop in C++. The *repeat* procedure is always implemented in the host language, which makes the connection between the three functions *repeat*, *cond* and *body* potentially difficult to identify in a Magnolia program.

```cpp
template <typename _Context, typename _State,
          class _body, class _cond>
struct while_ops {
  typedef _State State;
  typedef _Context Context;

  _body body;
  _cond cond;

  inline void repeat(State &state, const Context &context) {
    while (while_ops::cond(state, context)) {
      while_ops::body(state, context);
    }
  }
};
```

Listing 16 A problem with constraining concepts.

```
concept Graph = {
  type Graph;
  type Vertex;
  type VertexCollection;

  function adjacentVertices(g: Graph, v: Vertex)
    : VertexCollection;
  function vertices(g: Graph): VertexCollection;
  // predicate member(v: Vertex, vc: VertexCollection);
  // axiom adjacentVerticesAreVertices(
  //     v1: Vertex, v2: Vertex, g: Graph) {
  //   assert member(v2, adjacentVertices(g, v1)) =>
  //          member(v2, vertices(g))
  // }
}
```

## References

[1] Roland Carl Backhouse, Patrik Jansson, Johan Jeuring, and Lambert G. L. T. Meertens. "Generic Programming: An Introduction". In: *Revised Lectures of the Third International Spring School on Advanced Functional Programming*. Volume 1608. Lecture Notes in Computer Science. Springer-Verlag, 1998, pages 28–115. ISBN: 3-540-66241-3. DOI: 10.1007/10704973_2.

[2] Anya Helene Bagge. "Constructs & Concepts: Language Design for Flexibility and Reliability". [Last accessed 30-Sep-2022]. PhD thesis. PB 7803, 5020 Bergen, Norway: Research School in Information and Communication Technology, Department of Informatics, University of Bergen, Norway, 2009. ISBN: 978-82-308-0887-0. URL: http://www.ii.uib.no/~anya/phd/.

[3] Anya Helene Bagge. "Facts, Resources and the IDE/Compiler Mind-Meld". In: *Proceedings of the 4th International Workshop on Academic Software Development Tools and Techniques (WASDeTT'13)*. [Last accessed 30-Sep-2022]. Montpellier, France, July 2013. URL: http://wasdett.org/2013/submissions/wasdett2013_submission_10.pdf.

[4] Anya Helene Bagge, Valentin David, and Magne Haveraaen. "Testing with Axioms in C++ 2011". In: *Journal of Object Technology* 10 (2011), 10:1–32. ISSN: 1660-1769. DOI: 10.5381/jot.2011.10.1.a10.

[5] Anya Helene Bagge and Magne Haveraaen. "Interfacing Concepts: Why Declaration Style Shouldn't Matter". In: *Proceedings of the Ninth Workshop on Language Descriptions, Tools and Applications (LDTA '09)*. Edited by Torbjörn Ekman and Jurgen J. Vinju. Volume 253. York, UK: Elsevier, 2010, pages 37–50. DOI: 10.1016/j.entcs.2010.08.030.

[6] Anya Helene Bagge and Magne Haveraaen. "Specification of Generic APIs, or: Why Algebraic May Be Better Than Pre/Post". In: *Proceedings of the 2014 ACM*

*SIGAda Annual Conference on High Integrity Language Technology*. HILT '14. Portland, Oregon, USA: ACM, 2014, pages 71–80. ISBN: 978-1-4503-3217-0. DOI: 10.1145/2663171.2663183.

[7] Toby Bailey and Michael B. Gale. "Chesskell: A Two-Player Game at the Type Level". In: *Proceedings of the 14th ACM SIGPLAN International Symposium on Haskell*. Haskell 2021. Virtual, Republic of Korea: ACM, 2021, pages 110–121. ISBN: 9781450386159. DOI: 10.1145/3471874.3472987.

[8] Mike Barnett, K. Rustan M. Leino, and Wolfram Schulte. "The Spec# Programming System: An Overview". In: *Proceedings of the 2004 International Conference on Construction and Analysis of Safe, Secure, and Interoperable Smart Devices*. CASSIS'04. Marseille, France: Springer-Verlag, 2004, pages 49–69. ISBN: 3540242872. DOI: 10.1007/978-3-540-30569-9_3.

[9] Jan A. Bergstra and John V. Tucker. "Algebraic specifications of computable and semicomputable data types". In: *Theoretical Computer Science* 50.2 (1987), pages 137–181. ISSN: 0304-3975. DOI: 10.1016/0304-3975(87)90123-X.

[10] Michel Bidoit and Peter D. Mosses. *CASL User Manual - Introduction to Using the Common Algebraic Specification Language*. Volume 2900. Lecture Notes in Computer Science. Springer, 2004. ISBN: 3-540-20766-X. DOI: 10.1007/b11968.

[11] Luca Cardelli and Peter Wegner. "On Understanding Types, Data Abstraction, and Polymorphism". In: *ACM Comput. Surv.* 17.4 (Dec. 1985), pages 471–523. ISSN: 0360-0300. DOI: 10.1145/6041.6042.

[12] Jacques Carette, Russell O'Connor, and Yasmine Sharoda. *Building on the Diamonds between Theories: Theory Presentation Combinators*. 2019. DOI: 10.48550/arXiv.1812.08079.

[13] Manuel M. T. Chakravarty, Gabriele Keller, and Simon Peyton Jones. "Associated Type Synonyms". In: *Proceedings of the Tenth ACM SIGPLAN International Conference on Functional Programming*. ICFP '05. Tallinn, Estonia: ACM, 2005, pages 241–253. ISBN: 1595930647. DOI: 10.1145/1086365.1086397.

[14] Manuel M. T. Chakravarty, Gabriele Keller, Simon Peyton Jones, and Simon Marlow. "Associated Types with Class". In: *Proceedings of the 32nd ACM SIGPLAN-SIGACT Symposium on Principles of Programming Languages*. POPL '05. Long Beach, California, USA: ACM, 2005, pages 1–13. ISBN: 158113830X. DOI: 10.1145/1040305.1040306.

[15] Craig Chambers and the Cecil Group. *The Cecil language: specification and rationale, Version 3.2*. [Last accessed 30-Sep-2022]. 2004. URL: https://projectsweb.cs.washington.edu/research/projects/cecil/www/Release/doc-cecil-lang/cecil-spec.pdf.

[16] Benjamin Chetioui. *magnoliac: A Magnolia Compiler*. Dec. 2020. DOI: 10.5281/zenodo.6572953.

[17]  Benjamin Chetioui, Marius Larnøy, Jaakko Järvi, Magne Haveraaen, and Lenore Mullin. "P$^3$ Problem and Magnolia Language: Specializing Array Computations for Emerging Architectures". In: *Frontiers in Computer Science* 4 (2022). DOI: 10.3389/fcomp.2022.931312.

[18]  Koen Claessen and John Hughes. "QuickCheck: A Lightweight Tool for Random Testing of Haskell Programs". In: *Proceedings of the Fifth ACM SIGPLAN International Conference on Functional Programming*. ICFP '00. New York, NY, USA: ACM, 2000, pages 268–279. ISBN: 1581132026. DOI: 10.1145/351240.351266.

[19]  Manuel Clavel, Francisco Durán, Steven Eker, Santiago Escobar, Patrick Lincoln, Narciso Martí-Oliet, José Meseguer, Rubén Rubio, and Carolyn Talcott. *Maude Manual (Version 3.2.1)*. Feb. 2022.

[20]  Manuel Clavel, Francisco Durán, Steven Eker, Patrick Lincoln, Narciso Martí-Oliet, José Meseguer, and José F. Quesada. "Maude: specification and programming in rewriting logic". In: *Theoretical Computer Science* 285.2 (2002). Rewriting Logic and its Applications, pages 187–243. ISSN: 0304-3975. DOI: 10.1016/S0304-3975(01)00359-0.

[21]  Manuel Clavel, Steven Eker, Patrick Lincoln, and José Meseguer. "Principles of Maude". In: *Electronic Notes in Theoretical Computer Science* 4 (1996). RWLW96, First International Workshop on Rewriting Logic and its Applications, pages 65–89. ISSN: 1571-0661. DOI: 10.1016/S1571-0661(04)00034-9.

[22]  Thomas H. Cormen, Charles E. Leiserson, Ronald L. Rivest, and Clifford Stein. *Introduction to Algorithms, Third Edition*. 3rd. The MIT Press, 2009. ISBN: 0262033844.

[23]  James C. Dehnert and Alexander A. Stepanov. "Fundamentals of Generic Programming". In: [58], pages 1–11. ISBN: 3540410902. DOI: 10.1007/3-540-39953-4_1.

[24]  Razvan Diaconescu and Kokichi Futatsugi. *CafeOBJ report: The language, proof techniques, and methodologies for object-oriented algebraic specification*. Volume 6. World Scientific, 1998. ISBN: 978-9810235130.

[25]  Manuel Fähndrich, Michael Barnett, and Francesco Logozzo. "Embedded Contract Languages". In: *Proceedings of the 2010 ACM Symposium on Applied Computing*. SAC '10. Sierre, Switzerland: ACM, 2010, pages 2103–2110. ISBN: 9781605586397. DOI: 10.1145/1774088.1774531.

[26]  Robert W. Floyd. "Assigning Meanings to Programs". In: *Program Verification: Fundamental Issues in Computer Science*. Edited by Timothy R. Colburn, James H. Fetzer, and Terry L. Rankin. Dordrecht: Springer Netherlands, 1993, pages 65–81. ISBN: 978-94-011-1793-7. DOI: 10.1007/978-94-011-1793-7_4.

[27]  Tim Freeman and Frank Pfenning. "Refinement Types for ML". In: *Proceedings of the ACM SIGPLAN 1991 Conference on Programming Language Design and Implementation*. PLDI '91. Toronto, Ontario, Canada: ACM, 1991, pages 268–277. ISBN: 0897914287. DOI: 10.1145/113445.113468.

[28] Kokichi Futatsugi, Joseph A. Goguen, Jean-Pierre Jouannaud, and José Meseguer. "Principles of OBJ2". In: *Proceedings of the 12th ACM SIGACT-SIGPLAN Symposium on Principles of Programming Languages*. POPL '85. New Orleans, Louisiana, USA: ACM, 1985, pages 52–66. ISBN: 0897911474. DOI: 10.1145/318593.318610.

[29] Kokichi Futatsugi, Joseph A. Goguen, José Meseguer, and Koji Okada. "Parameterized Programming in OBJ2". In: *Proceedings of the 9th International Conference on Software Engineering*. ICSE '87. Monterey, California, USA: IEEE Computer Society Press, 1987, pages 51–60. ISBN: 0897912160.

[30] Ronald Garcia, Jaakko Järvi, Andrew Lumsdaine, Jeremy Siek, and Jeremiah Willcock. "An extended comparative study of language support for generic programming". In: *Journal of Functional Programming* 17.2 (2007), pages 145–205. DOI: 10.1017/S0956796806006198.

[31] Jeremy Gibbons. "Datatype-Generic Programming". In: *Proceedings of the 2006 International Conference on Datatype-Generic Programming*. SSDGP'06. Nottingham, UK: Springer-Verlag, 2006, pages 1–71. ISBN: 3540767851. DOI: 10.1007/978-3-540-76786-2_1.

[32] Joseph A. Goguen. "Higher-Order Functions Considered Unnecessary for Higher-Order Programming". In: *Research Topics in Functional Programming*. USA: Addison-Wesley Longman Publishing Co., Inc., 1990, pages 309–351. ISBN: 0201172364.

[33] Joseph A. Goguen. "Parameterized programming". In: *IEEE Transactions on Software engineering* 5 (1984), pages 528–543.

[34] Joseph A. Goguen. "Tossing algebraic flowers down the great divide". In: Springer, New York, 1999, pages 93–129. ISBN: 9789814021135.

[35] Joseph A. Goguen. "Types as Theories". In: *Topology and Category Theory in Computer Science*. USA: Oxford University Press, Inc., 1991, pages 357–385. ISBN: 0198537603.

[36] Joseph A. Goguen and Rod M. Burstall. "Introducing institutions". In: *Logics of Programs*. Edited by Edmund Clarke and Dexter Kozen. Berlin, Heidelberg: Springer Berlin Heidelberg, 1984, pages 221–256. ISBN: 978-3-540-38775-6. DOI: 10.1007/3-540-12896-4_366.

[37] Joseph A. Goguen, Claude Kirchner, Hélène Kirchner, Aristide Mégrelis, José Meseguer, and Timothy Winkler. "An Introduction to OBJ 3". In: *1st International Workshop on Conditional Term Rewriting Systems*. Orsay, France: Springer-Verlag, 1988, pages 258–263. ISBN: 3540192425. DOI: 10.1007/3-540-19242-5_22.

[38] Joseph A. Goguen and William Tracz. "An implementation-oriented semantics for module composition". In: *Foundations of Component-based Systems*. 2000, pages 231–263. ISBN: 0-521-77164-1.

[39] Joseph A. Goguen, Timothy Winkler, José Meseguer, Kokichi Futatsugi, and Jean-Pierre Jouannaud. "Introducing OBJ". In: *Software Engineering with OBJ: Algebraic Specification in Action*. Edited by Joseph Goguen and Grant Malcolm. Boston, MA: Springer US, 2000, pages 3–167. ISBN: 978-1-4757-6541-0. DOI: 10.1007/978-1-4757-6541-0_1.

[40] Andy Gordon and Cédric Fournet. *Principles and Applications of Refinement Types*. Technical report MSR-TR-2009-147. [Last accessed 30-Sep-2022]. Oct. 2009. URL: https://www.microsoft.com/en-us/research/publication/principles-and-applications-of-refinement-types/.

[41] Douglas Gregor and Jaakko Järvi. "Variadic Templates for C++0x". In: *Journal of Object Technology* 7.2 (Feb. 2008). Edited by Davide Ancona and Mirko Viroli. OOPS Track at the 22nd ACM Symposium on Applied Computing, SAC 2007, pages 31–51. ISSN: 1660-1769. DOI: 10.5381/jot.2008.7.2.a2.

[42] Douglas Gregor, Jaakko Järvi, Jeremy Siek, Bjarne Stroustrup, Gabriel Dos Reis, and Andrew Lumsdaine. "Concepts: linguistic support for generic programming in C++". In: *OOPSLA '06: Proceedings of the 21st annual ACM SIGPLAN conference on Object-oriented programming systems, languages, and applications*. Portland, Oregon, USA: ACM Press, 2006, pages 291–310. ISBN: 1-59593-348-4. DOI: 10.1145/1167473.1167499.

[43] Robert Griesemer, Raymond Hu, Wen Kokke, Julien Lange, Ian Lance Taylor, Bernardo Toninho, Philip Wadler, and Nobuko Yoshida. "Featherweight Go". In: *Proceedings of the ACM on Programming Languages* 4.OOPSLA (Nov. 2020). DOI: 10.1145/3428217.

[44] John L. Gustafson and Isaac T. Yonemoto. "Beating Floating Point at its Own Game: Posit Arithmetic". In: *Supercomputing Frontiers and Innovations* 4.2 (Apr. 2017), pages 71–86. DOI: 10.14529/jsfi170206.

[45] Hans-Christian Hamre. "Automated Verifications for Magnolia Satisfactions". Master's thesis. The University of Bergen, 2022.

[46] *Haskell Applicative => Monad Proposal*. [Last accessed 22-October-2021]. 2014. URL: https://wiki.haskell.org/Functor-Applicative-Monad_Proposal.

[47] Magne Haveraaen. "Axiom Based Testing for Fun and Pedagogy". In: *Formal Methods – Fun for Everybody*. Edited by Antonio Cerone and Markus Roggenbach. Cham: Springer International Publishing, 2021, pages 27–57. ISBN: 978-3-030-71374-4. DOI: 10.1007/978-3-030-71374-4_2.

[48] Magne Haveraaen, Jaakko Järvi, and Damian Rouson. *Reflecting on Generics for Fortran*. Technical report. [Last accessed 30-Sep-2022]. 2019. URL: https://j3-fortran.org/doc/year/19/19-188.pdf.

[49] Magne Haveraaen and Karl Trygve Kalleberg. "JAxT and JDI: The Simplicity of JUnit Applied to Axioms and Data Invariants". In: *OOPSLA Companion '08: Companion to the 23rd ACM SIGPLAN conference on Object-oriented programming systems languages and applications*. Nashville, TN, USA: ACM, 2008, pages 731–732. ISBN: 978-1-60558-220-7. DOI: 10.1145/1449814.1449834.

[50] Magne Haveraaen and Markus Roggenbach. "Specifying with syntactic theory functors". In: *Journal of Logical and Algebraic Methods in Programming* 113 (2020), page 100543. ISSN: 2352-2208. DOI: 10.1016/j.jlamp.2020.100543.

[51] Magne Haveraaen and Eric G. Wagner. "Guarded Algebras: Disguising Partiality so You Won't Know Whether Its There". In: *Recent Trends in Algebraic Development Techniques*. Edited by Didier Bert, Christine Choppy, and Peter D. Mosses. Berlin, Heidelberg: Springer Berlin Heidelberg, 2000, pages 182–200. ISBN: 978-3-540-44616-3. DOI: 10.1007/978-3-540-44616-3_11.

[52] Ralf Hinze and Johan Jeuring. "Generic Haskell: Practice and Theory". In: *Generic Programming: Advanced Lectures*. Edited by Roland Carl Backhouse and Jeremy Gibbons. Volume 2793. Lecture Notes in Computer Science. Springer, 2003, pages 1–56. ISBN: 978-3-540-45191-4. DOI: 10.1007/978-3-540-45191-4_1.

[53] Ralf Hinze and Johan Jeuring. "Weaving a web". In: *Journal of Functional Programming* 11.6 (2001), pages 681–689. DOI: 10.1017/S0956796801004129.

[54] Charles Antony Richard Hoare. "An axiomatic basis for computer programming". In: *Communications of the ACM* 12.10 (1969), pages 576–580.

[55] Paul Holser. *junit-quickcheck: Property-based testing, JUnit-style*. [Last accessed 30-May-2022]. 2014. URL: https://pholser.github.io/junit-quickcheck/site/1.0/.

[56] Atsushi Igarashi, Benjamin C. Pierce, and Philip Wadler. "Featherweight Java: A Minimal Core Calculus for Java and GJ". In: *ACM Transactions on Programming Languages and Systems* 23.3 (May 2001), pages 396–450. ISSN: 0164-0925. DOI: 10.1145/503502.503505.

[57] Jaakko Järvi, Jeremiah Willcock, and Andrew Lumsdaine. "Associated Types and Constraint Propagation for Mainstream Object-Oriented Generics". In: *Proceedings of the 20th Annual ACM SIGPLAN Conference on Object-Oriented Programming, Systems, Languages, and Applications*. OOPSLA '05. San Diego, CA, USA: ACM, 2005, pages 1–19. ISBN: 1595930310. DOI: 10.1145/1094811.1094813.

[58] Mehdi Jazayeri, Ruediger Loos, and David Musser. *Generic Programming: International Seminar on Generic Programming Dagstuhl Castle, Germany, April 27–May 1, 1998 Selected Papers*. Jan. 2000. ISBN: 978-3-540-41090-4. DOI: 10.1007/3-540-39953-4.

[59] Wolfram Kahl and Jan Scheffczyk. "Named Instances for Haskell Type Classes". In: *Proceedings of the 2001 Haskell Workshop*. Edited by Ralf Hinze. Volume 59. 2. Firenze, Italy, 2001.

[60] Deepak Kapur, David R. Musser, and Alexander A. Stepanov. "Operators and Algebraic Structures". In: *Proceedings of the 1981 Conference on Functional Programming Languages and Computer Architecture*. FPCA '81. Portsmouth, New Hampshire, USA: ACM, 1981, pages 59–64. ISBN: 0897910605. DOI: 10.1145/800223.806763.

[61] Oleg Kiselyov. *Functions with the variable number of (variously typed) arguments*. [Last accessed 24-May-2022]. June 2004. URL: https://okmij.org/ftp/Haskell/polyvariadic.html#polyvar-fn.

[62] Oleg Kiselyov, Ralf Lämmel, and Keean Schupke. *HList: Heterogeneous lists*. Version v0.5.2.0. [Last accessed 24-May-2022]. Feb. 2022. URL: https://hackage.haskell.org/package/HList-0.5.2.0/.

[63] Oleg Kiselyov, Ralf Lämmel, and Keean Schupke. "Strongly Typed Heterogeneous Collections". In: *Proceedings of the 2004 ACM SIGPLAN Workshop on Haskell*. Haskell '04. Snowbird, Utah, USA: ACM, 2004, pages 96–107. ISBN: 1581138504. DOI: 10.1145/1017472.1017488.

[64] Gary T. Leavens and Yoonsik Cheon. *Design by Contract with JML*. [Last accessed 20-September-2022]. 2006. URL: https://www.cs.ucf.edu/~leavens/JML/jmldbc.pdf.

[65] Simon Marlow, editor. *Haskell 2010 language report*. [Last accessed 22-October-2021]. 2010. URL: https://www.haskell.org/onlinereport/haskell2010/.

[66] Bertrand Meyer. "Applying 'design by contract'". In: *Computer* 25.10 (1992), pages 40–51. DOI: 10.1109/2.161279.

[67] Bertrand Meyer. *Eiffel: The Language*. Prentice Hall, New York, NY, 1991. ISBN: 0-13-247925-7.

[68] Andrey Mokhov. "United Monoids". In: *The Art, Science, and Engineering of Programming* 6.3 (Feb. 2022). DOI: 10.22152/programming-journal.org/2022/6/12.

[69] David R. Musser and Alexander A. Stepanov. "Generic Programming". In: *Symbolic and Algebraic Computation, International Symposium ISSAC'88, Rome, Italy, July 4-8, 1988, Proceedings*. Edited by Patrizia M. Gianni. Volume 358. Lecture Notes in Computer Science. Springer, 1988, pages 13–25. DOI: 10.1007/3-540-51084-2_2.

[70] Matt Noonan. "Ghosts of Departed Proofs (Functional Pearl)". In: *Proceedings of the 11th ACM SIGPLAN International Symposium on Haskell*. Haskell 2018. St. Louis, MO, USA: ACM, 2018, pages 119–131. ISBN: 9781450358354. DOI: 10.1145/3242744.3242755.

[71] Bruno C.d.S. Oliveira, Adriaan Moors, and Martin Odersky. "Type Classes as Objects and Implicits". In: *Proceedings of the ACM International Conference on Object Oriented Programming Systems Languages and Applications*. OOPSLA '10. Reno/Tahoe, Nevada, USA: ACM, 2010, pages 341–360. ISBN: 9781450302036. DOI: 10.1145/1869459.1869489.

[72] Peter Csaba Ölveczky. *Designing Reliable Distributed Systems - A Formal Methods Approach Based on Executable Modeling in Maude*. Undergraduate Topics in Computer Science. Springer, 2017. ISBN: 978-1-4471-6686-3. DOI: 10.1007/978-1-4471-6687-0.

[73] OpenMP Architecture Review Board. *OpenMP Application Program Interface Version 5.1*. [Last accessed 31-January-2022]. Nov. 2020. URL: https://www.openmp.org/wp-content/uploads/OpenMP-API-Specification-5-1.pdf.

[74]  Simon Peyton Jones, Andrew Tolmach, and Tony Hoare. "Playing by the rules: rewriting as a practical optimisation technique in GHC". In: *Proceedings of the 2001 Haskell Workshop*. Edited by Ralf Hinze. Volume 59. 2. Firenze, Italy, 2001.

[75]  Gabriel Radanne. *Typing Tricks: Diff lists*. [Last accessed 24-May-2022]. Aug. 2016. URL: https://drup.github.io/2016/08/02/difflists/.

[76]  Patrick M. Rondon, Ming Kawaguci, and Ranjit Jhala. "Liquid Types". In: *Proceedings of the 29th ACM SIGPLAN Conference on Programming Language Design and Implementation*. PLDI '08. Tucson, AZ, USA: ACM, 2008, pages 159–169. ISBN: 9781595938602. DOI: 10.1145/1375581.1375602.

[77]  Donald Sannella and Andrzej Tarlecki. "Extended ML: An Institution-Independent Framework for Formal Program Development". In: *Category Theory and Computer Programming: Tutorial and Workshop, Guildford, U.K. September 16–20, 1985 Proceedings*. Edited by David Pitt, Samson Abramsky, Axel Poigné, and David Rydeheard. Springer Berlin Heidelberg, 1986, pages 364–389. ISBN: 978-3-540-47213-1. DOI: 10.1007/3-540-17162-2_133.

[78]  Donald Sannella and Andrzej Tarlecki. "Mind the Gap! Abstract Versus Concrete Models of Specifications". In: *Mathematical Foundations of Computer Science 1996, 21st International Symposium, MFCS'96, Cracow, Poland, September 2-6, 1996, Proceedings*. Edited by Wojciech Penczek and Andrzej Szalas. Volume 1113. Lecture Notes in Computer Science. Springer, 1996, pages 114–134. DOI: 10.1007/3-540-61550-4_143.

[79]  Gabriel Scherer. *The 6 parameters of ('a, 'b, 'c, 'd, 'e, 'f) format6*. [Last accessed 24-May-2022]. Apr. 2014. URL: http://gallium.inria.fr/blog/format6/.

[80]  Gabriel Scherer and Didier Rémy. "GADTs Meet Subtyping". In: *Programming Languages and Systems*. Edited by Matthias Felleisen and Philippa Gardner. Berlin, Heidelberg: Springer Berlin Heidelberg, 2013, pages 554–573. ISBN: 978-3-642-37036-6. DOI: 10.1007/978-3-642-37036-6_30.

[81]  Tom Schrijvers, Simon Peyton Jones, Manuel Chakravarty, and Martin Sulzmann. "Type Checking with Open Type Functions". In: *Proceedings of the 13th ACM SIGPLAN International Conference on Functional Programming*. ICFP '08. Victoria, BC, Canada: ACM, 2008, pages 51–62. ISBN: 9781595939197. DOI: 10.1145/1411204.1411215.

[82]  Jeremy Siek, Lie-Quan Lee, and Andrew Lumsdaine. *The Boost Graph Library: User Guide and Reference Manual*. Boston, MA, USA: Addison-Wesley Longman Publishing Co., Inc., 2002. ISBN: 0-201-72914-8.

[83]  Jeremy Siek and Andrew Lumsdaine. "Concept checking: Binding parametric polymorphism in C++". In: *Proceedings of the First Workshop on C++ Template Programming*. [Last accessed 01-Oct-2022]. Erfurt, Germany, Oct. 2000, page 12. URL: https://citeseerx.ist.psu.edu/viewdoc/summary?doi=10.1.1.22.427.

[84]  Jeremy G. Siek. "A Language for Generic Programming". PhD thesis. USA, 2005. ISBN: 0542308096.

[85] Jeremy G. Siek. "The C++0x "Concepts" Effort". In: *Generic and Indexed Programming: International Spring School, SSGIP 2010, Oxford, UK, March 22-26, 2010, Revised Lectures*. Edited by Jeremy Gibbons. Berlin, Heidelberg: Springer Berlin Heidelberg, 2012, pages 175–216. ISBN: 978-3-642-32202-0. DOI: 10.1007/978-3-642-32202-0_4.

[86] Jeremy G. Siek and Andrew Lumsdaine. "A language for generic programming in the large". In: *Science of Computer Programming* 76.5 (2011). Special Issue on Generative Programming and Component Engineering (Selected Papers from GPCE 2004/2005), pages 423–465. ISSN: 0167-6423. DOI: 10.1016/j.scico.2008.09.009.

[87] Alexander Stepanov and Paul McJones. *Elements of Programming*. 1st. Addison-Wesley Professional, 2009. ISBN: 032163537X.

[88] Andrew Sutton. "Overloading with Concepts". In: *Overload* 24 (Dec. 2016). [Last accessed 30-Sep-2022]. URL: https://accu.org/journals/overload/24/136/sutton_2316/.

[89] Andrew Sutton and Jonathan I. Maletic. "Emulating C++0x concepts". In: *Science of Computer Programming* 78.9 (2013), pages 1449–1469. ISSN: 0167-6423. DOI: 10.1016/j.scico.2012.10.009.

[90] Andrew Sutton and Bjarne Stroustrup. "Design of Concept Libraries for C++". In: *Software Language Engineering*. Edited by Anthony Sloane and Uwe Aßmann. Berlin, Heidelberg: Springer Berlin Heidelberg, 2012, pages 97–118. ISBN: 978-3-642-28830-2. DOI: 10.1007/978-3-642-28830-2_6.

[91] Benoît Vaugon. *A new format implementation based on GADTs*. [Last accessed 24-May-2022]. May 2013. URL: https://github.com/ocaml/ocaml/issues/6017.

[92] Dana N. Xu. "Hybrid Contract Checking via Symbolic Simplification". In: *Proceedings of the ACM SIGPLAN 2012 Workshop on Partial Evaluation and Program Manipulation*. PEPM '12. Philadelphia, Pennsylvania, USA: ACM, 2012, pages 107–116. ISBN: 9781450311182. DOI: 10.1145/2103746.2103767.

[93] Dana N. Xu, Simon Peyton Jones, and Koen Claessen. "Static Contract Checking for Haskell". In: *Proceedings of the 36th Annual ACM SIGPLAN-SIGACT Symposium on Principles of Programming Languages*. POPL '09. Savannah, GA, USA: ACM, 2009, pages 41–52. ISBN: 9781605583792. DOI: 10.1145/1480881.1480889.

[94] Edward Z. Yang. *Type classes: confluence, coherence and global uniqueness*. [Last accessed 01-February-2022]. 2014. URL: http://blog.ezyang.com/2014/07/type-classes-confluence-coherence-global-uniqueness/.

**About the authors**

**Benjamin Chetioui** is the corresponding author for this paper. Contact Benjamin at benjamin.chetioui@uib.no.

**Jaakko Järvi** is a professor of Software Engineering at the University of Turku. Contact Jaakko at jaakko.jarvi@utu.fi.

**Magne Haveraaen** is a professor in Informatics at the University of Bergen and the head of Bergen Language Design Laboratory (BLDL). Contact Magne at magne.haveraaen@uib.no.